\documentclass[letterpaper,twocolumn,10pt]{article}
\usepackage{usenix,epsfig}

\newcommand{\comment}[1]{}
\newcommand{\safe}{\textsf{SAFE}}
\newcommand\keyword{\normalfont}
\usepackage{color}

\usepackage{courier}

\usepackage[utf8]{inputenc}
\usepackage{listings}

\usepackage{graphicx}
\usepackage{url}
\usepackage{upquote}
\usepackage{color}
\usepackage{enumitem}
\usepackage{subcaption}

\usepackage[utf8]{inputenc}
\usepackage{listings}

\begin{document}

\date{}

\title{\Large \bf SAFE: A Declarative Trust Management System with Linked Credentials}

\author{
Vamsi Thummala, Jeff Chase \\
Duke University\\
\{vamsi, chase\}@cs.duke.edu
}

\maketitle

\subsection*{Abstract}

We present \keyword{SAFE}, an integrated system for managing trust using a logic-based
declarative language. Logical trust systems authorize each request by
constructing a proof from a context---a set of authenticated logic statements
representing credentials and policies issued by various principals in a
networked system.

A key barrier to practical use of logical trust systems is the problem of
managing proof contexts: identifying, validating, and assembling the
credentials and policies that are relevant to each trust decision. This paper
describes a new approach to managing proof contexts using context linking and
caching. Credentials and policies are stored as certified logic sets named by
secure identifiers in a shared key-value store. \keyword{SAFE} offers language
constructs to build and modify logic sets, link sets to form unions, pass them
by reference, and add them to proof contexts. \keyword{SAFE} fetches and
validates credential sets on demand and caches them in the authorizer. We
evaluate and discuss our experience using \keyword{SAFE} to build secure
services based on case studies drawn from practice: a secure name service
resolver, a secure proxy shim for a key value store, and an authorization
module for a networked infrastructure-as-a-service system with a federated
trust structure.

\section{Introduction}
\label{sec:intro}
\begin{figure*}[t!]
    \centering 
    \includegraphics[width=6.6in,height=3.5in]{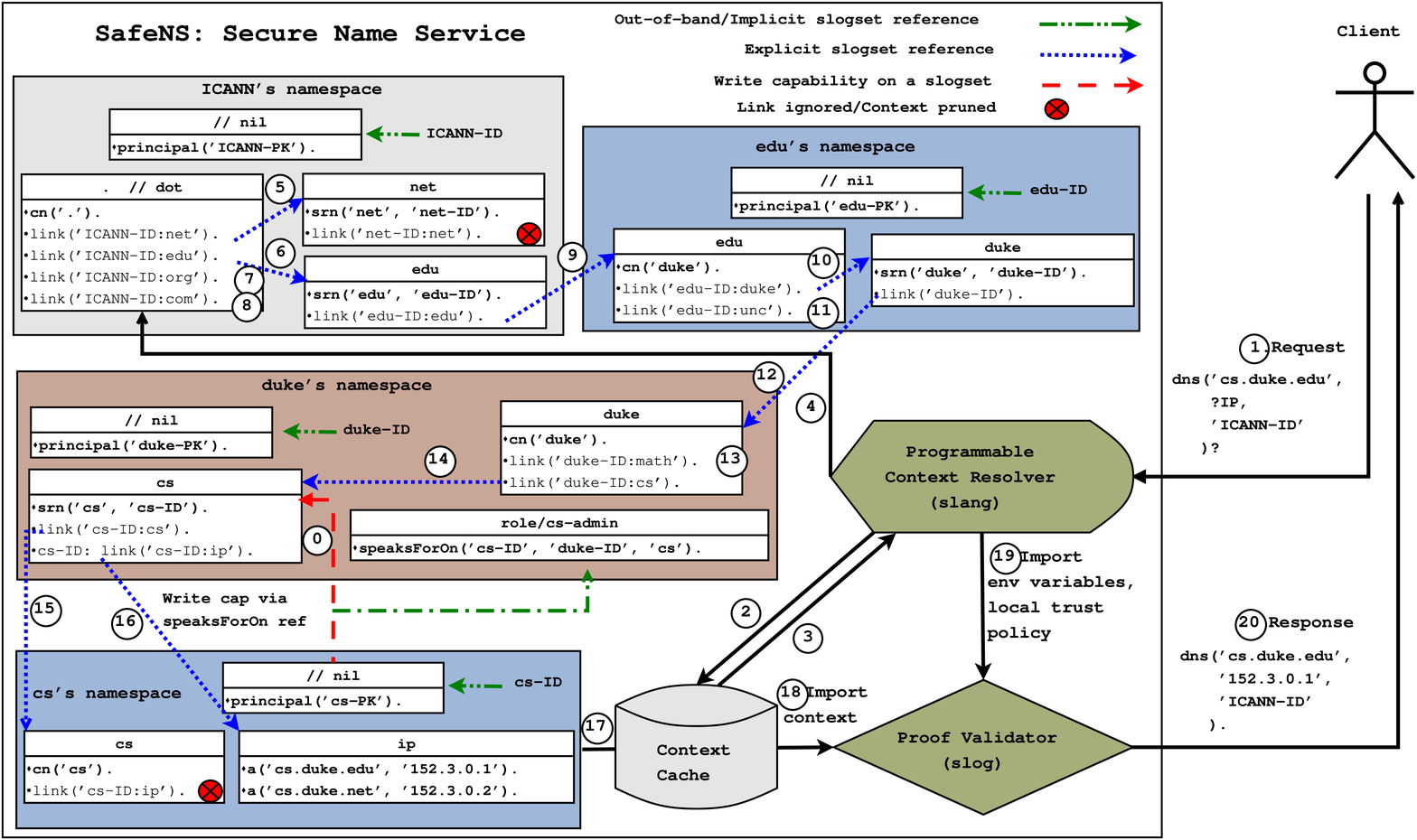}
    \caption{\small Workflow illustrating credential discovery, context building,
    context caching, and proof validation process for a secure name
    service---SafeNS---emulating DNSSec resolver implemented in \keyword{SAFE}.
    The credentials are issued a priori (step 0) and materialized as slogsets
    in a shared distributed store (SafeSets). The principal {\tt cs} writes to
    a slogset owned by {\tt duke} due to a {\tt speaksForOn} capability issued
    by the owning principal (step 0). The \keyword{SAFE} process starts with a
    bearer slogset reference ({\tt ICANN-ID}) provided by the client and
    traverses the credential graph via links---building a tailored context as
    per the resolver's programming logic written in \keyword{slang} (steps
    1-17). The \keyword{slang} runtime invokes the \keyword{slog} interpreter
    importing the relevant context---and \keyword{slog} interpreter validates
    the proof based on local trust anchors and policies, and certifies the
    response (steps 18-20) similar to certified validation in
    SD3~\cite{jim:sd3:2001}.}
    \label{fig:safedns} 
\end{figure*} 

Trust management deals with specifying and interpreting security policies,
credentials, and relationships among entities in a system to reach an
authorization decision~\cite{blaze96:trust}. Authorization determines whether a
certain request (e.g., read, write) on one or more {\it objects} (e.g., file,
process), is permitted by the requesting {\it principal}. Credentials are
statements about the principals issued by concerned parties delegating the
chain of trust. To enforce a given policy, each entity in the trust management
system has a reference monitor (a guard) that uses credentials in conjunction
with the request to infer the authorization decision.

Over time, the formal foundations of trust management systems have converged on
logic-based declarative languages---{\it trust logic}.  One prominent early
example of trust management is SPKI/SDSI~\cite{spki-rfc2693}, in which
participants exchange statements binding principals to names in local name
spaces. Halpern {\it et al.} showed that the naming language in SPKI/SDSI has a
logical semantics~\cite{halpern2001logic} and Howell {\it et al} provided
formal semantics~\cite{Howell:2000:FSS:867818} that has roots in logic.  Li
{\it et al.}~\cite{Li03:constraint, abac-spki} showed that SPKI/SDSI naming
maps to a Role-based Trust (RT) language, and that the RT language in turn
reduce to datalog with constraints~\cite{ceri89-datalog-dare}, a logic language
with well-understood formal properties including tractability. 

However despite the flexibility and extensibility of trust logics, their
application to practice is limited. We observe some key obstacles in harnessing
the power of trust logics in practical distributed systems:

\begin{enumerate}
\setlength\itemsep{0em}
\item \emph{Credential discovery.} Given a query, how to identify and assemble the
tailored proof context that is relevant to an authorization decision?  
\item \emph{Credential freshness and revocation.} Scalable revocation is a major issue
with the deployment of PKI based systems. The challenge is how to revoke an issued
credential and propagate the changes in a timely fashion in a distributed
setting?  
\item \emph{Usability.} Usability is an important goal for making trust logics
practical and approachable. How to make trust logics programmable so that
symbolic names are used rather than low-level identifiers for principals and
objects?  How to specify trust policies and perform access checks so that the
system integrates naturally with the service programming environment?
\item \emph{Federated trust.} How to name and collaboratively share resources
among federated trust domains adhering to local trust policies and system
constraints?
\end{enumerate}

To this end we built an integrated logical trust system called
\keyword{SAFE}.\footnote{\keyword{SAFE} is an acronym for Secure Authorization
for Federated Environments. The source repository is available
at~\cite{safe-repo}.}  At the core of \keyword{SAFE} is a simple trust logic
(\keyword{SAFE} logic or \keyword{slog}) based on extended datalog with
constraints. \keyword{Slog} is similar in spirit to
Binder~\cite{detreville:binder:2002}, SD3~\cite{jim:sd3:2001},
Soutei~\cite{Pimlott06:soutei}, and SENDLog~\cite{Sendlog:Abadi:2007}.  These
logics can capture important examples of secure network systems.  For example,
SD3 has been used to implement DNSSec~\cite{jim:sd3:2001} resolver, and SENDLog
has been used to implement secure routing protocols that build on the
declarative routing approach.

What is novel about \keyword{SAFE} is its approach to managing {\it
contexts}---sets of credentials and other logic content---and the relationships
among them. \keyword{SAFE} builds on a key concept called {\it set linking}, a
powerful technique to organize sets of logic statements. A {\it link} is a
meta-predicate of the credential set and serves much like a hyperlink in HTML
documents.  A logic set (\keyword{slogset}) may link to a target set by its
name: the link incorporates the target as a subset, forming a union.  A
construction procedure integrates set linking with common primitives for
delegation and endorsement, naturally materializing a graph in which each set
links to the other sets needed to substantiate it. A guard specifies a context
as a linked union of top-level sub-contexts (e.g., \keyword{slogsets}
associated with the subject, object, and policy). The transitive closure of the
sub-context contains all sets relevant to a given authorization decision.  Set
linking also facilitates flexible policy because it easy to attach policy rule
sets to nodes in the graph. The cost of linking is low because common subsets
are cached at the authorizer. A further advantage with linking and shared
storage is that the credentials can be prefetched and cached naturally to
support the materialized context sets for the future queries.
Figure~\ref{fig:safedns} shows a workflow of SafeNS---a secure name service
emulating the DNSSec resolver implemented in SAFE (see Sec~\ref{sec:safedns}).

\comment{
For example, Bob may pass a set link as a reference to Charlie, which may be
passed on to Alice when requesting for access. Alice can use the reference
passed by Charlie to fetch the requisite credential sets and infer the access
privilege---or use the local policy to grant access at her discretion. SafeSets
enables such hybrid policies with flexible querying and credential retrieval
either on-demand or a priori at the discretion of the authorizer.
\comment{Moreover, in future when Bob decides to revoke access to Charlie, he
can simply publish an updated set by retracting the statement that Charlie is
his coworker. Bob also controls the refresh time on his published credentials,
and hence any concerned parties including Alice will notice the updated
credential on its expiry or subsequent fetch on the same set identifier,
limiting Charlie's access privileges. Unlike capability based systems where the
issuer needs to track the subject, set linking and in situ update of
credentials through a shared store in SAFE makes revocation scalable in
practice.}
}

In addition, \keyword{SAFE} offers a scripting language called
``\keyword{slang}'' that manipulates sets of authenticated logic statements
(\keyword{slogsets}) as first class content objects with unique names.
Contrast to slog, which is used as a certifying proof engine, \keyword{slang}
is used primarily for credential discovery, certificate issuing and revocation,
and tailoring proof context based on authorizer's policies. Separating the
credential discovery process from proof validation is important to ensure the
inference is tractable. We are also motivated to design \keyword{slang} to make
trust logics more usable in practice---\keyword{slang} is declarative and makes
the crypto operations transparent and provides built-in hooks to integrate with
the service programming environment.

\keyword{Slang} provides support for higher order logical semantics such as
{\tt speaksFor} and {\tt aggregation}, which cannot be captured in
\keyword{slog} due to lack of of function symbols and nesting.  \keyword{Slang}
offers language constructs to build and modify \keyword{slogsets}, publish them
by name, fetch them by reference, and add them as sub-contexts to a proof
context. The name space enables \keyword{slogsets} to be passed by reference,
fetched on demand, and cached after validation at the receiver. In future
exchanges the receiver may retrieve sets from its cache as needed, avoiding the
need to transmit and validate them again.  \keyword{Slogsets} are themselves
stored in a shared, secure, and distributed credential store called as
\emph{SafeSets}.  Each object in SafeSets is signed by its speaker to enable a
third-party to verify the authenticity and integrity of the source. The shared
store is also a basis for addressing perennial problems with PKI certificate
management, e.g., revocation, renewal, and key rotation.

{\bf Contributions.} Our research focuses on practical challenges for using
logical trust in secure networked systems. Our premise is that trust logic can
be as fast as PCAs and as simple identity-based access control schemes (e.g.,
ACLs) in the common case, while enabling rich and flexible declarative trust
with a precise and rigorous logical semantics and verifiable policies. More
generally, we believe that logical trust can be a fundamental enabler for a
network security architecture that is richer, safer, and more flexible than the
architecture in place today (e.g., X.509 and CA hierarchy). We make these contributions: 
\begin{enumerate}
\setlength\itemsep{0em}
\item A high productivity programming tool for managing credentials
declaratively based on language extensions to trust logics including tractable
delegation with {\tt speaksFor}, policy mobility, and server
integration (Sec~\ref{sec:approach}).
\item A shared, secure, and distributed credential store that leverages
the concept of set linking to organize credentials (Sec~\ref{sec:impl}).
\item An evaluation to demonstrate that \keyword{SAFE} is practical including
the implementation of secure name service (SafeNS) and an authorization system
GENI~\footnote{GENI is a networked infrastructure-as-a-service (IaaS) with
autonomous IaaS providers linked in a federated trust structure.} (SafeGENI)
in a hundred lines of \keyword{SAFE} scripting language (Sec~\ref{sec:eval}).
\end{enumerate}

\comment{
\begin{table*}
 \centering
 \begin{tabular}{p{2cm} r r r r r r r} \\ \hline
      & PolicyMaker/KeyNote & SPKI/SDSI & Snowflake & PCA & RT & SecPAL \\ \hline \hline
    Formal Foundation & Graph search & Graph search & Tuple reduction & Higher order logic & $Datalog_{C}$ & $Datalog_{C}$ \\ \hline
    Authorization Procedure & Graph search & Graph search & Tuple reduction & Higher order logic & $Datalog_{C}$ & $Datalog_{C}$ \\ \hline
    Authorization Complexity & Graph search & Graph search & Tuple reduction & Higher order logic & $Datalog_{C}$ & $Datalog_{C}$ \\ \hline
    Distributed Credential Discovery & Graph search & Graph search & Tuple reduction & Higher order logic & $Datalog_{C}$ & $Datalog_{C}$ \\ \hline
    Certificate Encoding & Graph search & Graph search & Tuple reduction & Higher order logic & $Datalog_{C}$ & $Datalog_{C}$ \\ \hline
    Delegation of Authority (SpeaksFor) & Graph search & Graph search & Tuple reduction & Higher order logic & $Datalog_{C}$ & $Datalog_{C}$ \\ \hline
  \end{tabular}
\caption{\small Comparison of Trust Management Systems}
\label{tab:trust-survey}
\end{table*}
}

\section{Logical Trust on the Network}
\label{sec:design}
In this section, we review the elements of trust management
systems: how to name principals, objects, and logic sets; the trust
requirements to satisfy in a networked system; the security assumptions and
design choices that guide the implementation of \keyword{SAFE}. Some of the
design choices are influenced by previous work on trust
management---specifically, the local namespaces of
SPKI/SDSI~\cite{spki-rfc2693} and self-certifying names
in~\cite{SFS:SOSP:1999,Mazieres:1998}. However one major difference of \keyword{SAFE}
compared to SPKI/SDSI is set linking: \keyword{SAFE} provides an {\it explicit}
linking capability at the level of sets of logic statements
(\keyword{slogsets}) rather than implicit linking on names as in SPKI/SDSI.
Explicit linking makes credential discovery~\cite{clarke01} and revocation
practical and scalable compared to SPKI/SDSI: the name resolution requires the
authorizer to resolve the relevant certificates among a potentially large set
of certificates in the \emph{right} order; in the common case, the authorizer
may simply act as a compliance proof checker putting the onus on the requester
to carry the trust relationships.


\subsection{Naming}

\keyword{SAFE} relies on cryptographic keys for identifying the principals
following SPKI/SDSI. A principal is a self-signed public key---i.e., a
principal possesses the private key corresponding to the public key that is
signed---or an attested public key by a trusted third party following the
current Certifying Authorities (CAs) model. Every principal speaks indirectly
through sub-principals by creating and assigning roles or by issuing {\tt
speaksFor} delegation to alleviate key rotation issues and keeping the master
keypair stored securely off-line.  Principals create local namespaces to keep
track of resources they own, capabilities they receive, endorsements they make,
or bookmark references to other principal's namespace that may contain relevant
trust policies for a later retrieval. 

A principal's namespace is identified globally and uniquely by a pair
{\textless {\tt principal}, {\tt name}\textgreater} known as a \emph{self-certifying
identifier} or scid for short. Self-certifying names provide the useful
property that any entity in a distributed system can verify the binding between
a corresponding public key and the local name without relying on a trusted
third party~\cite{Mazieres:1998}. Self-certifying names thus provide a
decentralized form of data origin authentication. Without loss of generality,
the scids are defined as $H_1$({\tt principal}):$H_2$({\tt name}), where $H_1$
and $H_2$ are cryptographic hash corresponding to the tuples, giving us a fixed
length scid irrespective of the key sizes and symbolic names. 

In \keyword{SAFE}, a principal's namespace corresponds to a \keyword{slogset}
in which credentials and policies are stated as logical statements. In addition
to scid, each \keyword{slogset} can be identified by a secure reference (or id
for short), which is formed by taking a hash of scid, i.e., $H_3$($H_1$({\tt
principal}):$H_2$({\tt name})). An identity \keyword{slogset} is a special set
without a name and contains principal's public key. For the identity
\keyword{slogset}, the scid and the id are equal.

Objects in \keyword{SAFE} are identified by their scids. \keyword{SAFE}
recognizes three types of objects: credential objects (\keyword{slogsets}) for
which the local name is chosen by the issuer; resource objects for which the
local name is auto-generated using an RFC 4122 GUID/UID; and content objects
for which the local name is the hash of the contents. \keyword{SAFE} provides a
built-in {\tt rootId()} to extract the controlling principal name from an
object name.

\keyword{SAFE} does not mandate a single global namespace or a central
certifying authority. In \keyword{SAFE}, each principal is a certifying
authority. Explicit endorsement and the ability to link \keyword{slogsets} by
reference provides more flexible design choices without assuming any naming
convention a priori. Where required, the hierarchical naming can be represented
by aggregating {\tt scids} explicitly. For example, a DNS request such as {\tt
cs.duke.edu} can be represented in \keyword{SAFE} as:

\noindent
{\small
$H_1(P_{\tt .}):H_2({\tt edu})).H_1(P_{\tt edu}):H_2({\tt duke}).H_1(P_{\tt duke}):H_2({\tt cs})$,
}
where $P_{\tt name}$
represents the principal owning the name and $H_1, H_2$ are hash functions.
Once $P_{\tt .\{dot\}}$ is available (browsers can bootstrap trust anchors
following today's practices), \keyword{SAFE} can be queried to infer the IP
address securely subsuming DNSSec or other secure DNS implementations. Such
aggregation of scids across multiple principals (with `.' as the delimiter) to
form a secure compound names is known as \emph{safe resource naming} or SRN for
short. We explain the secure name resolution implemented in \keyword{SAFE}
further in Sec~\ref{sec:safedns}.

\subsection{Requirements}

To use the trust logic in a networked system the following requirements must be
met: (1) network messages can be authenticated as originating from a named
principal; (2) each statement in the logic is authenticated to its named
speaker; (3) each object name is securely bound to a given principal who
controls the name; (4) to the extent that one party accepts or relies on
another's statements, the parties must agree on the meaning of predicate
symbols and names used in those statements.

Without loss of generality we meet the first two requirements by taking
principal name constants as public keys (or hashes) and transporting statements
in certificates signed by the named speaker, following SDSI.

The third requirement is trivially met with local namespaces, i.e., each
principal hash its own object name space and requests for those objects are
served only through a server controlled by that principal.

The fourth requirement is met by standards and conventions in the code.
\keyword{SAFE} applications may define their own vocabulary of predicates.
Note that common conventions are needed only for interoperability, but not for
soundness. The soundness of \keyword{SAFE} inference requires only that
statements are authentic and that the relevant name constants are unique and
distinct.

\subsection{Assumptions}
We make the certain assumptions about the threat model, SafeSets availability,
principal keypairs, and credential linking.
\begin{itemize}
\item The \keyword{SAFE} client running the inference should be a part of
trusted computing base. All other components including SafeSets need not be
trusted.
\item SafeSets is configured to be highly available storage system. With
scalable key-value stores, this requirement is easily met.
\item Every principal creates sub-principals to speaks for them and stores the
master key-pair securely off-line.
\item All delegations, speaksFor, set construction, and linking are done {\it
explicitly} through logical assertions, i.e., \keyword{SAFE} does not support
implicit delegations as in~\cite{abadi93:accesscalculus}.
\end{itemize}

\section{Managing Credentials Declaratively}
\label{sec:approach}
This section presents an overview of how \keyword{SAFE} applications use trust
logic languages (\keyword{slog} and \keyword{slang}) to build and issue
credentials as \keyword{slogsets}, how \keyword{slogsets} can be linked for
credential discovery, and how a set linking supports policy mobility.  It
illustrates with examples from SafeGENI, which is described in a related
technical report~\cite{chase14:safegeni}.  The GENI trust architecture defines
several classes of authority services to manage user identity and authorize
user activity. These services are decentralized: each authority service may
have multiple instances, and the set of instances may change over time.  In
addition, users may delegate various rights to one another using a capability
model. SafeGENI specifies all of these structures using logic.

A notable feature of \keyword{SAFE} is integration of the trust logic with a
scripting layer that manipulates logic content and invokes the proof engine.
\keyword{Slog} is a tractable logic language that generates a proof locally
from a supplied proof context. Who supplies the proof context and how is it
assembled? The \keyword{slang} provides scripting constructs to build and
modify \keyword{slogsets} from templates, link them to form unions, publish/post them
(e.g., as certificates written to SafeSets), pass or fetch them by reference,
and add them to query contexts for trust decisions.

An application includes \keyword{slang} code to construct any logic content it
issues or fetch and cache logic content from other parties, assemble proof
contexts, issue trust queries, and organize its credentials and policies.  Each
participant in a networked system chooses the \keyword{slang} code that it
executes: the participants exchange declarative logic content, but not
scripting code. 

\subsection{SAFE Logic (slog)}
\label{sec:slog}

Slog is a elementary trust logic based on constrained datalog, which is a
subset of first-order logic. Logic statements in \keyword{slog} are written in
datalog augmented with the classic {\bf says} operator of BAN belief
logic~\cite{Burrows:1990} and ABLP logic~\cite{lampson92:authentication}.
Statements are built up from atomic formulas (atoms) and the logical operators
conjunction and implication.  An atom is a predicate symbol applied to a list
of parameters, which may be variables or constants representing principals,
objects, or values. Every atom has a first parameter representing a principal
who {\bf says} it (the {\it speaker}). Consider this slog statement:

\begin{verbatim}
authorize(?Subj) :- 
  Alice: coworker(?Subj).
\end{verbatim}

This statement reads {\it ``self infers authorize(?Subj), for any given subject
represented by a variable ?Subj, if the principal Alice says
coworker(?Subj) is true''}.   The {\tt :-} is datalog syntax for logical
implication: this statement is a {\it rule}.  The text to the left of the {\tt
:-} is the {\it head} of the rule, and the text to the right is the {\it body}.
The head is a single atom whose parameters may include one or more variables
({\tt ?Subj}).  A rule allows the checker to infer that the head is true,
for some substitution of its variables with constants, if the body is true
under that substitution.  The body is a sequence of atoms (called {\it goals})
separated by commas, which indicate conjunction: all of the atoms in the body
must be true for the rule to ``fire''. All variables in the head must also
appear in the body. A {\it fact} is a statement with no body, and therefore no
variables. The checker takes any fact in the context as true.  The predicate in
an atom or fact represents a property, attribute, role, relationship, right,
power, or permission associated with the principals and/or objects named in its
parameters.

Each atom is bound to a speaker. In the example, the atoms in the body are
prefixed with a {\bf says} operator ({\tt :}) naming the principal {\tt Alice}.
If an atom does not name a speaker then the default speaker is {\tt
\$Self}---the local authorizer who applies the statement. Note also that the
speaker of an atom in the body of a rule may be a variable. Consider this rule
from SafeGENI:

\begin{verbatim}
memberAuthority(?X) :- geniRoot(?Geni),
  ?Geni: memberAuthority(?X).
\end{verbatim}

This rule reads ``{\it self infers that ?X is a Member Authority, for any given
?X, if some principal ?Geni says it is, and self believes that ?Geni is the
GENI root}''. A GENI root is a principal that is accepted by members of a GENI
federation to endorse authority services and IaaS providers (aggregates).  The
policy may include or import a fact designating a {\tt geniRoot} trust anchor.
A GENI Member Authority is a principal that is authorized to issue statements
about user identity including roles, privileges, account status, and key
endorsements.  This rule states that the authorizer believes an assertion ({\tt
memberAuthority(?X)}) if it is spoken by any principal ({\tt ?Geni}) possessing
a certain attribute ({\tt geniRoot(?Geni)}).  This form of rule is known as an
{\it attribute-based delegation}~\cite{rt-abac}.


The goals in policy rules such as these capture the meaning of delegation of
trust. The delegation is restricted both by the speaker of the goal and the
predicate used.   For example, the rule above trusts a {\tt geniRoot} only to
endorse a {\tt memberAuthority}, and it trusts the endorsed principal only as a
{\tt memberAuthority}.  Other rules in SafeGENI delegate specific additional
powers to principals with the attributes {\tt geniRoot} or {\tt
memberAuthority}.


\begin{table}[t!]
  \small
  \begin{tabular}{p{2.6cm} p{4.6cm}} \hline 
  {\bf Function } & {\bf Description}\\ \hline \hline
  {\tt fetch(?SetRef)} & fetch a transitive closure of \keyword{slogset} ref by traversing all the links\\ \hline
  {\tt fetchSRN( ?SetRef, ?SRN)} & fetch a transitive closure of
  \keyword{slogset} ref by traversing the links as guiding by the safe resource
  name (SRN)\\ \hline
  {\tt post( ?SetContents)} & post the set contents and return the \keyword{slogset} reference.\\ \hline
  \end{tabular}
  \caption{\small \keyword{Slang} library functions implementing the SafeSets Client API.}
  \label{tab:slang-func}
\end{table}

\subsection{SAFE Language (slang)}
\label{sec:slang}

\keyword{Slang} is a simple hybrid functional-logic programming language with
an extended logic syntax supporting higher order structures with nested
function symbols. \keyword{Slang} is designed to be used as a scripting
language for credential discovery, credential pruning (tailoring proof context
based on authorizer's policies and the issued request), and certificate issuing
and revocation. A \keyword{slang} program is a set of logic statements similar
to the \keyword{slog} program but structurally akin to Prolog programs rather
than Datalog programs. However, a crucial difference is that \keyword{slang}
programs are local to the authorizer and \emph{not transported over network}
unlike \keyword{slogsets}. Alternatively, \keyword{slang} programs assist the
credential discovery process and building proof context tailored to the
request---but the programs themselves are not part of trust infrastructure and
inference. \keyword{SAFE} considers the authorization decision as valid only if
\keyword{slog} performs the inference.

A \keyword{slang} program permits usage of higher order constructs to process
collections: lists, nested predicates, and \keyword{slogsets} as \emph{first class}
objects. For example, \keyword{slogsets} can be manipulated as values directly by
assigning them to variables and passing them as arguments to other functions.

Other important distinction from \keyword{slog} is that \keyword{slang}
statements may act as functions that return values, including
\keyword{slogsets}. In general, \keyword{slang} programs execute as a
functional evaluation rather than inference: the evaluation follows a
deterministic path with no backtracking, presuming that for each slang
predicate there is at most one rule with a matching head (the common case).

The design of \keyword{slang} is motivated in part by our experience with
building authorization system for GENI.
\begin{itemize}

\item Current approaches for authorization are limited: either support a high
level language compromising on proof tractability (e.g.,
PCA~\cite{appel99:proof-carrying}, PolicyMaker~\cite{blaze96:trust},
KeyNote~\cite{blaze03:keynote}, NAL~\cite{schneider11:nal}) or restricted
language leaving the credential gathering to the applications (e.g.,
SecPAL~\cite{Becker10:secpal}, Soutei~\cite{Pimlott06:soutei}).

\item We used \keyword{slog} as an embedded language library from a generic
purpose language and observed the impedance mismatch between language layers.
For example, the \keyword{slog} program is passed as a string from the host
language, which results in deferring the ``safety'' properties of
\keyword{slog} until actual execution time.

\item We observed common patterns (fetching, publishing, and renewing) for
managing credentials which are handled efficiently using a high level
abstraction---manipulating \keyword{slogsets} as values.

\item Certain useful logical primitives such as {\tt speaksFor} and {\tt
aggregation} cannot be captured at the \keyword{slog} layer but can be easily
achieved at higher layer without compromising tractability of the logical
inference~\cite{Grohe:2010}.

\item Writing \keyword{slog} code directly is tedious and prone to mistakes
since the principals and objects are hashed values rather than simple mnemonic
names. \keyword{Slang} makes it particularly convenient to define policies
naturally through the use of lexically scoped program variables, environment
variables that capture system properties, and builtin library functions that
operate on \keyword{slogsets} directly.

\item \keyword{Slang} also supports programming through policy templates so
that policies are written once and instantiated accordingly as per the
environment context and scope.

\item Lastly, \keyword{slang} is declarative and resembles \keyword{slog} closely while
being expressive. \keyword{Slang} performs traditional scripting functions: file
manipulation, escaping to the host environment for program execution, and
variable substitution.
\end{itemize}

\begin{lstlisting}[float, showstringspaces=false, caption={Geni root endorses and IdP.}, label=code:geniRoot]

defcon endorse(?IdP) :-
  spec('endorse an identity provider'),
  ''endorse/idp/$IdP''{ // slogset name
    identityProvider($IdP). // variable subst
    link($Self). // link to self (geni) ID set
  } // end of slogset definition
end // end of slang function

definit ?Ref := endorse('IdP-ID'), post(?Ref).

\end{lstlisting}

Consider the GENI example in Code Listing~\ref{code:geniRoot}. Geni root
creates a \keyword{slogset} with a name {\tt ``endorse/idp/\$IdP''} and issues
simple \keyword{slog} statements endorsing an identity provider (IdP). The
statements enclosed within {\tt \{ \}} forms a first-class \keyword{slogset}
term extending the standard logic syntax to sets of statements. \keyword{Slang}
supports lexical scoping and global substitution of variables. {\tt \$IdP} is a
variable passed from \keyword{slang} to \keyword{slog}, which is
interpolated---substituted by its value---at runtime.  A \keyword{slang} rule
tagged with a {\tt def} keyword declares the rule as a function that returns
the value of the last atom on that rule.  The various slang rule types have
additional behaviors to integrate with the application and with SafeSets, and
to extend the scripting primitives. \keyword{Slang} predefines some functions
with prefixes {\tt defenv} for initializing environment variables; {\tt defcon} for
constructing slogsets; {\tt defguard} for entry points to \keyword{slang}
program for access checking incoming requests; and {\tt definit} for
bootstrapping the \keyword{slang} program. Other built-in functions that
implement the SafeSets client API are shown in Table~\ref{tab:slang-func}.
Code Listing~\ref{code:pi} shows how a Project Investigator (PI) relies on
local trust policy and bearer reference provided by the subject to determine
whether the subject is a valid geni user. In this case, it is the subject's
responsibility to provide a reference to a \keyword{slogset}, which contains a
statement that the issued by IdP that the subject is a geni user. The
authorizer augments the proof context constructed from bearer reference with
its own local policy and invokes the slog inference.  This example demonstrates
the flexible and extensible authorization in \keyword{SAFE} in which hybrid
policies are fetch on-demand or a priori at the discretion of the authorizer.
See the technical report for a complete working example of
GENI~\cite{chase14:safegeni}.

\keyword{Slang} also supports {\tt speaksFor} and {\tt speaksForOn} delegation
in a restricted form. Our implementation of {\tt speaksForOn} closely follows
restriction delegation proposed in Snowflake
project~\cite{Howell:2000:FSS:867818, Howell:2000}.  However, unlike
ABLP~\cite{abadi93:accesscalculus} and Snowflake~\cite{Howell:2000:FSS:867818}
projects, we do not view {\tt speaksFor} as a primitive form of delegation.
Recall that \keyword{slog} supports attribute-based delegation with {\bf says}
as the primitive operator. In \keyword{SAFE}, the restricted {\tt speaksFor} is
defined as follows: if a subject {\tt Alice} issues {\tt speaksFor} delegation
capability for {\tt Bob}, then {\tt Alice} grants {\tt Bob} to write to any
\keyword{slogset} owned by {\tt Alice}. Similarly, {\tt speaksForOn} restricts
the capability to a particular \keyword{slogset} named by {\tt Alice}. Now when
{\tt Bob} is issuing statements for {\tt Alice}, it is {\tt Bob's}
responsibility to write to an appropriate \keyword{slogset} under the {\tt
Alice} namespace. The {\tt speaksFor} and {\tt speaksForOn} are stated as
ordinary \keyword{slog} facts but interpreted specially by \keyword{slang} to
achieve the desired functionality. In \keyword{SAFE}, we use {\tt speaksForOn}
delegation for joint ownership among principals in a group and
principal/sub-principal roles, and delegating authority to a set of attributes
(\keyword{slogset}) collectively rather than specifying each attribute
individually as in ABAC~\cite{rt-abac}.

Lastly, \keyword{slang} also provides support for {\tt aggregation}, which
cannot be supported directly in \keyword{slog} without loosing
tractability~\cite{Grohe:2010}.  Aggregation is useful to
implement advanced features in trust logics such as threshold/manifold
structures as used by SPKI/SDSI.



\begin{lstlisting}[float, showstringspaces=false, caption={Project Investigator (PI) relies on local trust policy and bearer reference passed by the {\tt ?Subject} to determine whether {\tt ?Subject} is a valid geni user.}, label=code:pi]

defenv ?Geni :- 'geni-ID'. // hash-of-geni-PK

defcon trustStructure() :-
  spec('trust structure at the authorizer'),
  'policy/localTrust'{
     identityProvider(?X) :- 
       geniRoot(?Geni), 
       ?Geni: identityProvider(?X).
     geniUser(?U) :- 
       identityProvider(?IdP), 
       ?IdP: geniUser(?U).
   }
end

defguard isGeniUser(?Subject, ?BearerRef) :- 
  {
     import('policy/localTrust').
     import($BearerRef). // slogset reference
     geniRoot($Geni). // substitute env var
     geniUser($Subject)? // subst slang var
  }
end
\end{lstlisting}

\subsection{Set Linking and Support Sets}
\label{sec:linking}

The power of trust logics creates new obstacles to harnessing their power in
practical distributed systems. For example, authorization in decentralized
federated environments involve finding the necessary credentials that satisfy
the local policy for a given access control request. A key obstacle is {\it
credential discovery}: a trust decision may require reasoning from statements
drawn from various sources, requiring a method to discover and retrieve them.
In general, credential discovery is the process of finding the chain of
credentials that delegates the authority from the source to the requester.
Credential discovery is different from the certificate path discovery in X.509
certificates~\cite{Elley01:cert-path-disc} since credentials in trust
management systems generally have more complex meanings than simply binding
names to public keys. For example, a credential chain is often a DAG, rather
than a linear path as in X.509.

Most previous work in trust management assumes that authorizer has already
gathered all the potentially relevant credentials before a request is made and
does not consider credential discovery problem
further~\cite{blaze96:trust, blaze03:keynote, Becker10:secpal}.
Even if the authorizer gathers all the credentials a priori, a crucial issue is
that tailoring the credentials per query request rather than supplying all the
available credentials to the proof context---since the cost of inference
depends on the size of the proof context.

To make the credential discovery possible and efficient, we propose {\it set
linking} to build trust chains by linking relevant logic sets in advance.  A
further advantage of our approach is that it naturally supports caching of
context sets for future decisions.

The construction procedure is distributed across the participants who issue and
receive credentials---\keyword{slogsets} containing endorsements and delegations. Each
participant collects and stores the received credentials by using
meta-predicate {\tt link} to cross reference them into credential sets that it maintains.
The issuer of a credential uses {\tt link} in the set constructor to link the
new set to any of its own credential sets that support its authority to issue
the credential. Code Listing~\ref{code:geniRoot} shows {\tt link} predicate is
used as a reference to Geni root's ID set. If all issuers follow this convention
then by induction the transitive closure of any given credential contains the
totality of upstream credentials that an authorizer needs to validate it---the
credential's {\it support set}. In this way, set linking naturally forms
delegation chains in the credential graph. The authorizer uses its local
checker to validate that these chains lead back to one or more trust anchors
(e.g., {\tt geniRoot}) according to its policies.

The bearer reference link (id) provided by the subject ({\tt ?BearerRef} in
Code Listing~\ref{code:pi}) makes the authorizer to inspect the user
endorsements that she received. Linked support sets make it easy for an
authorizer to obtain all credentials necessary for an authorization decision by
``pulling'' a credential set token passed as an argument in a request, fetching
the closure of the linked subsets recursively, caching them, and adding them to
the proof context. Each participant is free to organize its credentials as it
sees fit, possibly across multiple sets. What is important is that each issuer
links sufficient support into each endorsement or delegation, and that each
requester passes sufficient support to justify each request. The linked sets
may contain a superset of what is required: the authorizer's \keyword{slog}
engine searches the context for relevant content. We emphasize that in
practice, a server finds frequently linked supporting credentials in its cache,
and does not fetch or validate them again on each request. 

In this way, set linking organizes credentials and policies into a DAG that
facilitates discovery and assembly of proof contexts. We note that fetch is
cycle-safe: it ignores any cycles, which do not affect the contents of the
closure.  The DAG is collaboratively editable: each node in the DAG is
controlled by its owners, and changes to a set by its owners are visible in
other sets that link to it. The sets are, in essence, materialized views for
standard queries, in which the subset owners control what statements to include
in the views.

Further, set linking naturally supports {\it policy mobility}: the guard
policies can be defined once by the trust anchors and the authorizers can use
them wherever applicable by simply fetching from SafeSets. 

\subsection{End-to-End Example: SafeNS}
\label{sec:safedns}

We implemented a secure name service---SafeNS---in \keyword{SAFE}. Using
SafeNS, we emulate the DNSSec resolver in \keyword{SAFE}.
Figure~\ref{fig:safedns} illustrative the end-to-end workflow of credential
discovery, set linking, context building and pruning using SafeNS as an
example. 

Given a name service request by the client, the browser/client-agent augments
the request with a bootstrapped reference to the root ({\tt ICAANN-ID})
\keyword{slogset} and passes to the SafeNS resolver. The SafeNS resolver uses
the bearer reference to initiate the credential discovery process using a
\keyword{slang} library function {\tt fetchSRN()}. The resolver fetches the
root set referenced by {\tt ICAANN-ID} and matches the common name for the root
({\tt cn(.)}) with the first name token {\tt `.'} by issuing a \keyword{slog}
query. If the \keyword{slog} query returns true---i.e., the safe resource name
(SRN) binds/matches with the \keyword{slogset} local name given by the {\tt
srn()} predicate---then the search continues further following the {\tt link}
predicates until a closure is reached. The \keyword{slang} runtime builds a
tailored context based on the SRN, and once the search completes,
\keyword{slang} invokes \keyword{slog} with the relevant proof context. The
\keyword{slog} process validates the proof based on local trust anchors ({\tt
ICAANN-ID}) and policies, and certifies the response. The workflow also
illustrates the use of {\tt speaksForOn} issued by the principal {\tt duke}
delegating ownership to the principal {\tt cs} on a particular set named {\tt
cs}.

SafeNS resolver curtails the proof context at each stage of the credential
discovery by using a constrained function {\tt fetchSRN()} rather than {\tt
fetch()} (see Table~\ref{tab:slang-func}). {\tt fetch()} fetches the
transitive closure name service endorsements starting from root to the proof
context, which is prohibitive if not curtailed to the relevant context. It is
important to note that {\tt fetchSRN()} is implemented as a \keyword{slang}
library function (in 30 lines of code) rather than a native implementation. The
context resolver is programmable: for example, the authorizer can add custom
rules to accept content only from authoritative servers located in the US
region.

\section{Implementation}
\label{sec:impl}
\begin{figure*} 
    \centering 
    \includegraphics[width=6.6in]{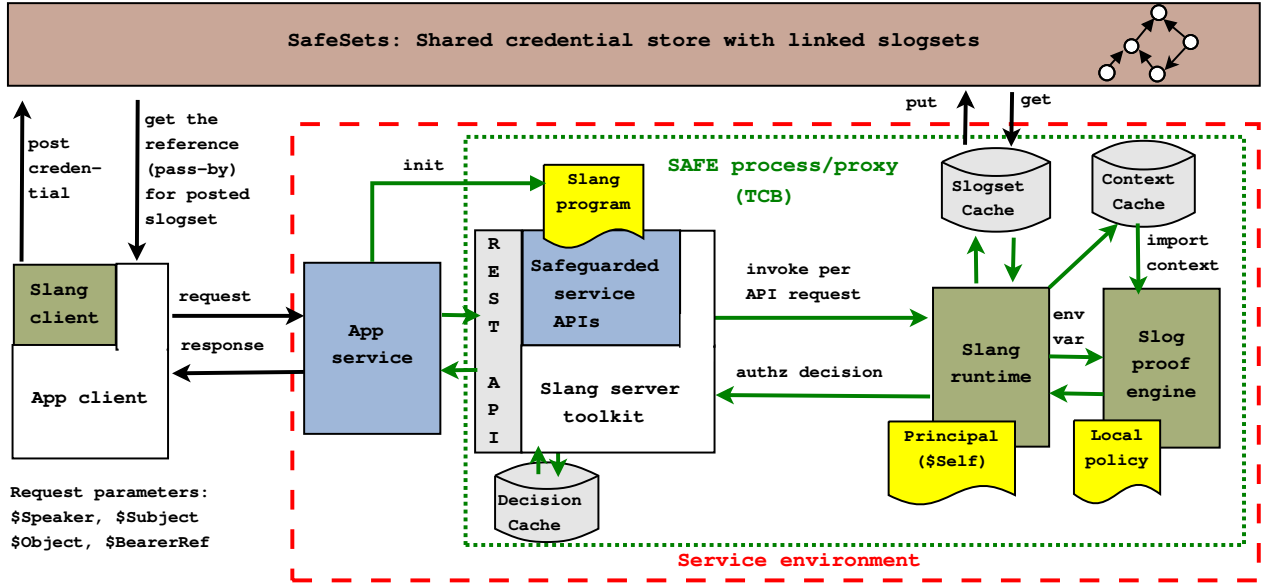}
    \caption{\small Server access control using \keyword{SAFE}. The \keyword{SAFE}
    instance runs as a separate process with a loaded program of
    \keyword{slang} that contains context building procedure, the principal's
    signing key ({\tt \$Self}), and the authorizer's local policies specified
    in \keyword{slog}. The server application installs \keyword{slang} code in
    the \keyword{SAFE} process, which registers all the {\tt defguard} APIs for
    access control checks. Credentials are passed as references to signed logic
    sets (slogsets) in a shared distributed store (SafeSets).  The
    \keyword{SAFE} process fetches slogsets on demand, validates the signature
    and speaker, and caches them for use by the \keyword{slog} interpreter.} 
    \label{fig:safe-arch} 
\end{figure*} 

The \keyword{SAFE} project builds on the earlier research in logic-based trust
management by focusing on logical trust as a systems problem. Elements of the
\safe\ project include integration with application service frameworks, a
deployment structure that facilitates cross-language interoperability,
programming tools to construct policies and credentials for logical trust, and
a decoupling from the external representations to transport the logic.

\subsection{SAFE Runtime}

\keyword{SAFE} runs as an interpreter with one or more \keyword{slang} programs
loaded into it. The \keyword{slang} code can run from command-line tools or
within a co-located \keyword{SAFE} process invoked through a REST API, or it
can integrate directly with JVM applications.  The code's behavior is
determined not just by the slang code itself but also by the logic content
passed to it. Any participant may add local rules to tailor the policies to
local needs, without changing the \keyword{slang} program. Participants may
even formulate rules and exchange them over the network as the system executes.  

The interpreter is stateless, so participants may restart it and/or reload
slang programs at any time: it affects only the access control for future
requests. Slang programs are composable: it is easy to add code to customize
the local behavior. Changing the program leaves other software and state
unchanged at the site.   

\keyword{SAFE} is implemented in Scala language including the inference engine
written from scratch. Source code for \keyword{SAFE} is available
at~\cite{safe-repo}.

\keyword{Slang} and SafeSets offer an integrated solution for sharing
authenticated logic sets in a networked system. Each authorizer's local
\keyword{SAFE} runtime interacts with the SafeSets service to support the set
abstractions of slang by fetching referenced sets on demand, caching their
logic content, and assuring the freshness and validity of logic content passed
to the proof engine. The \keyword{SAFE} runtime handles secure slogset id
generation, post, fetch, and cryptographic operations automatically and
transparently.

Because the slang scripting language abstracts these details and hides them
from applications, SafeSets is a replaceable component within the
\keyword{SAFE} architecture. However, the idea of using a shared decentralized
certificate store generalizes to other models for storing and authenticating
the sets.  In principle, logic sets could be stored in secure web directories
maintained for the owning principals, or in some scenarios might be stored bare
in a trusted metadata service, e.g., for use of trust logic within a single
service provider domain.

When a program defines a \keyword{slogset} (using {\tt defcon)}), the builtin
encoder consumes the meta-facts and encodes the information they contain into
the selected certificate format.  When \keyword{SAFE} fetches a certificate,
the builtin decoder validates the certificate, extracts the contents, and
materializes it as an in-memory \keyword{slogset}.  The \keyword{slogset}
represents the relevant meta-information from the certificate as logic
meta-facts.  These facts are available to the slog inference engine if the set
is added to the context for a query.

The SafeSets service itself is implemented as a proxy shim to a scalable Riak
key/value store~\cite{riak}.  On a post operation, the slogset id serves as the
key, and the named certificate is the value.  The shim checks access for post
operations: it verifies that the value (a certificate containing a logic set)
is signed under a public key whose hash yields the slogset id, when when hashed
with the local name. The shim is implemented using \keyword{SAFE} itself: it is
a \keyword{SAFE} process with \keyword{slang} code that invokes ordinary
certificate parsing and validation, queries the meta-attributes, and performs
the guard check.  SafeSets clients access the Riak store only through the shim,
which serves the Riak request protocol.  This is a simple example of using
\keyword{SAFE} to ``safeguard'' a network service transparently, as an
alternative to modifying the service or integrating with a service framework.

\subsection{SafeSets Certificate Store}
\label{sec:safesets}

\keyword{Slang} and SafeSets offer an integrated solution for sharing
authenticated logic sets in a networked system. Each authorizer's local
\keyword{SAFE} runtime interacts with the SafeSets service to support the set
abstractions of slang by fetching referenced sets on demand, caching their
logic content, and assuring the freshness and validity of logic content passed
to the proof engine. The \keyword{SAFE} runtime handles secure slogset id
generation, post, fetch, and cryptographic operations automatically and
transparently.

Because the slang scripting language abstracts these details and hides them
from applications, SafeSets is a replaceable component within the
\keyword{SAFE} architecture. However, the idea of using a shared decentralized
certificate store generalizes to other models for storing and authenticating
the sets.  In principle, logic sets could be stored in secure web directories
maintained for the owning principals, or in some scenarios might be stored bare
in a trusted metadata service, e.g., for use of trust logic within a single
service provider domain.

When a program defines a \keyword{slogset} (using {\tt defcon)}), the builtin
encoder consumes the meta-facts and encodes the information they contain into
the selected certificate format.  When \keyword{SAFE} fetches a certificate,
the builtin decoder validates the certificate, extracts the contents, and
materializes it as an in-memory \keyword{slogset}.  The \keyword{slogset}
represents the relevant meta-information from the certificate as logic
meta-facts.  These facts are available to the slog inference engine if the set
is added to the context for a query.

\keyword{SAFE} supports compact, reliable encoding of logic sets in X.509
certificates (using a string encoding within an attribute field) and also in a
native SAFE format.  The crypto layer represents all semantic content of any
signed certificate internally in common logic, including builtin predicates for
meta-attributes such as expiration time, encoding type, and so on.   It
generates the encoded cert from a \keyword{slogset} containing the required
meta-attributes as facts, which are easy to specify directly in \keyword{slang}
set constructors ({\tt defcon}. The native SAFE cert format is not subject to
the arbitrary length constraints of X.509 certificates, and also improves
compactness by hashing public keys embedded as principal names in the logic.
All of our experiments use the native \keyword{SAFE} cert format.

The SafeSets service itself is implemented as a proxy shim to a scalable Riak
key/value store~\cite{riak}.  On a post operation, the slogset id serves as the
key, and the named certificate is the value.  The shim checks access for post
operations: it verifies that the value (a certificate containing a logic set)
is signed under a public key whose hash yields the slogset id, when when hashed
with the local name. The shim is implemented using \keyword{SAFE} itself: it is
a \keyword{SAFE} process with \keyword{slang} code that invokes ordinary
certificate parsing and validation, queries the meta-attributes, and performs
the guard check.  SafeSets clients access the Riak store only through the shim,
which serves the Riak request protocol.  This is a simple example of using
\keyword{SAFE} to ``safeguard'' a network service transparently, as an
alternative to modifying the service or integrating with a service framework.

\subsection{Server Integration}
\label{sec:server}

Application server frameworks can use \keyword{SAFE} as a proxy or invoke
through REST API to check access control for client operations on the objects
they server (see Figure~\ref{fig:safe-arch}). Suppose that each API method of
the service has registered a corresponding guard in the slang program through
{\tt defguard}.  When a request enters the Web application or service
framework, it invokes SAFE to evaluate a declared guard whose name matches the
requested method, passing a list of variables named in the request.
\keyword{SAFE} evaluates the guard and returns the result to the service
framework, which rejects the service request if the result is false.  Ideally
there is no change to the application itself, other than defining
\keyword{slang} guards for each method. The \keyword{SAFE} runtime passes the
request parameters for each method to the registered \keyword{SAFE} guards via
{\tt defguard}.

\begin{table*}
 \centering
 \begin{tabular}{p{2cm} r r r r r r r} \\ \hline
    function & compute hash & verify signature & sign a set &  parse a set & null inference & fetch a set & post a set \\ \hline \hline
    \parbox[t]{2cm}{latency (ms)} & 0.13 & 0.56 & 12.14 & 2.2 & 0.09 & 7.8 & 27.4 \\ \hline
  \end{tabular}
\caption{\small Micro-benchmarks of basic operations in \keyword{SAFE} on 1kB of
payload per certificate. Fetch and post costs are network latencies over WAN
for reading and writing/updating a \keyword{slogset} to SafeSets. Keys are
2048-bit RSA keys. Hash function is set to SHA-256. The \emph{null} inference
is the minimum latency penalty for querying \keyword{slog} through
\keyword{slang}.}
\label{tab:ops}
\end{table*}

\subsection{Fetching, Validation, and Caching}
\keyword{SAFE} fetches \keyword{slogsets} automatically on first reference to a
token in the \keyword{slang} program. The client side code performs a
cycle-safe recursive fetch and the requests are parallelized using a thread
pool for reduced latency.  After each subset is fetched, \keyword{SAFE}
validates the signature, parses the certificate, and authenticates the stated
speaker of each statement.  If the certificate is valid, its contents are
extracted into an in-memory logic set, including meta-attributes. Sets created
in slang or imported from SafeSets are cached in an in-memory as {\it slogset
cache}. The fetch checks the set cache for each subset token encountered during
the recursive fetch.


When a sub-context is assembled and dispatched to the inference engine,
\keyword{SAFE} {\it renders} it to a internal context format and caches it in a
{\it context cache}. A rendered context set is flattened, then indexed to speed
up the inference engine, which must search for rules with heads matching each
goal. The expiry date on the context cache is set to the \emph{lowest} expiry
dates from the collections of sub-contexts that are assembled.  \keyword{SAFE}
tracks the period of validity internally to expunge any expired content from
the caches.  A leaf subset expires from the \keyword{slogset} cache at the
expiration time of its containing certificate. We also added support to
invalidate a cached context at the earliest expiration time of any statement it
contains, so the proof engine sees only fresh logic content. If an expired
certificate is reissued, then a fetch pulls the fresh certificate from the
store automatically.


We enhanced server integration by adding initial support for a query {\it
decision cache} on the Web server side. The result cache optimizes repeated
operations by a given subject on a given object, by avoiding the inference
check entirely for a configurable time. We do not report results from decision
cache. 

\subsection{Certificate Management}
\label{sec:slang-example}

\keyword{SAFE} and SafeSets address a number of longstanding challenges for
certificate management.

{\bf Renewal.}  An issuer may renew an expired certificate by posting the
renewed certificate to SafeSets with the same identifier, overwriting the
expired certificate. If a \keyword{SAFE} authorizer encounters an expired set
it uses the identifier to fetch a new version automatically, and retries the
query.

{\bf Revocation.}  An issuer may change a posted logic set at any time or
``poison'' the set to invalidate it. Of course, a change or poison does not
take effect if an authorizer uses an old copy of the set from its cache. An
issuer may control the expiration times to bound the time that a set remains in
an authorizer's cache by setting the {\tt refresh} time on the published set.
An authorizer may refresh sets in its cache at its discretion, even if they
have not expired. 

{\bf Rotation.}  \keyword{SAFE} with SafeSets names principals by their public
key hashes. If a principal loses or rotates its key-pair, then sets that
incorporate the stale key in their set identifiers must be regenerated. To
avoid potentially expensive set identifiers regeneration, \keyword{SAFE}
advocates each principal to create sub-principals which {\tt speaksFor} the
master on a given role (e.g., signing, encryption).

The \keyword{SAFE} approach to managing credentials also raises some potential
concerns which we discuss in Sec~\ref{sec:discussion}.

\begin{figure}
  \begin{center}
    \scalebox{.7}{\input{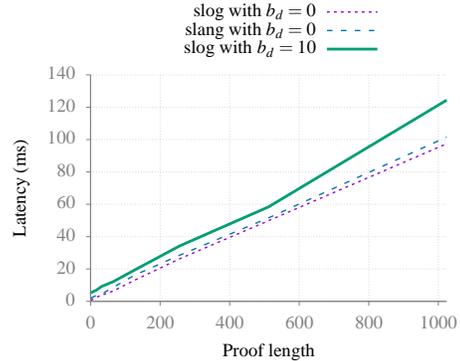}}
  \end{center}
  \caption{Cost of inference with varying proof length and degree of
backtracking $b_d$. The latency measurements show that the inference cost
scales linearly if $b_d$ is kept low. The plot also shows the overhead of
calling \keyword{slog} from \keyword{slang} program is minimal ($<5$\%). }
  \label{fig:proofLength}
\end{figure}

\section{Evaluation}
\label{sec:eval}
\begin{figure*}[t!]
\centering
\begin{subfigure}{.31\textwidth}
  \centering
   \scalebox{.6}{\input{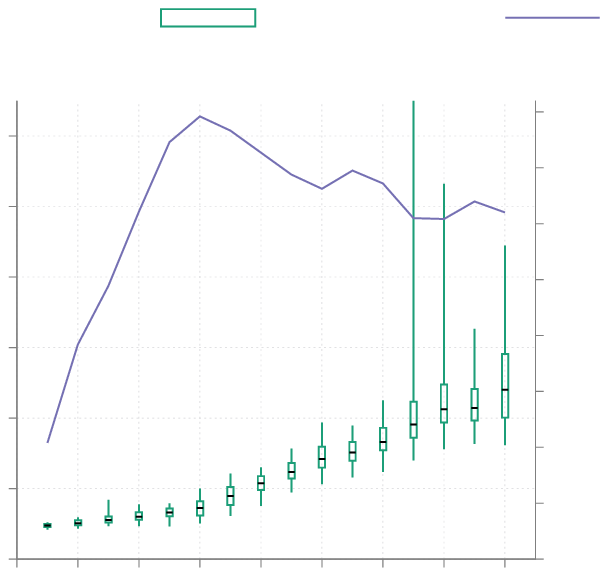}}
  \caption{\small SafeSets write without authorization}
  \label{fig:safesets-direct-post}
\end{subfigure}%
~
\begin{subfigure}{.31\textwidth}
  \centering
   \scalebox{.6}{\input{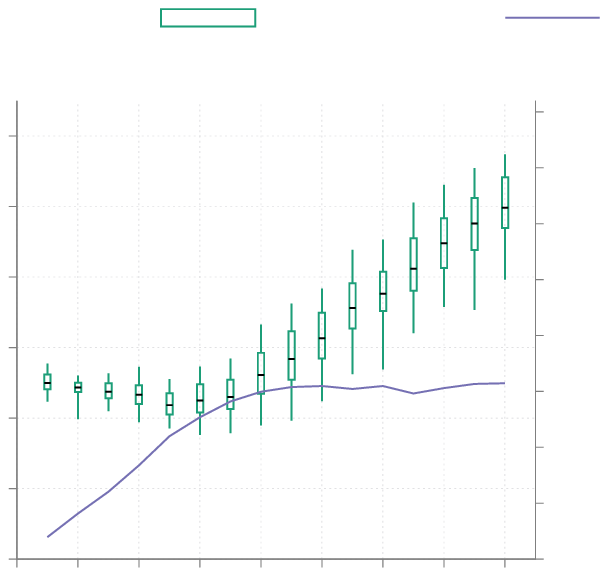}}
  \caption{\small SafeSets write without authorization but parsed on the server (baseline)}
  \label{fig:safesets-withoutVerify-post}
\end{subfigure}%
~
\begin{subfigure}{.31\textwidth}
  \centering
   \scalebox{.6}{\input{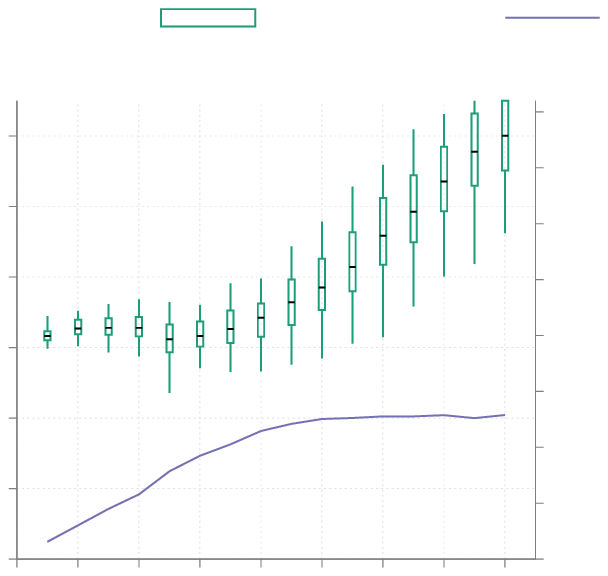}}
  \caption{\small SafeSets write with \keyword{SAFE} authorization enabled}
  \label{fig:safesets-authorized-post}
\end{subfigure}%
\caption{\small Performance comparison of issuing a \emph{write} to SafeSets store (a
write is a post with \keyword{slogset} signing excluded).}
\label{fig:safesets-perf}
\end{figure*}

\begin{figure*}[t!]
\centering
\begin{subfigure}{.48\textwidth}
  \centering
   \scalebox{.7}{\input{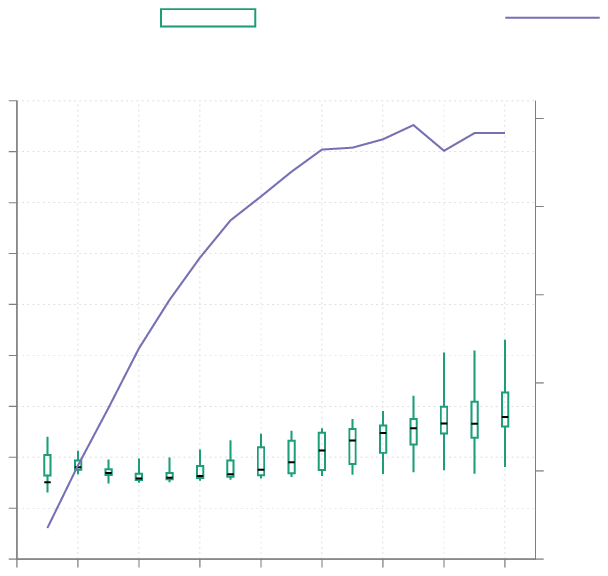}}
  \caption{\small SafeNS proof validation cost for delegations of length 4}
  \label{fig:safedns-perf}
\end{subfigure}%
~
\begin{subfigure}{.48\textwidth}
  \centering
   \scalebox{.7}{\input{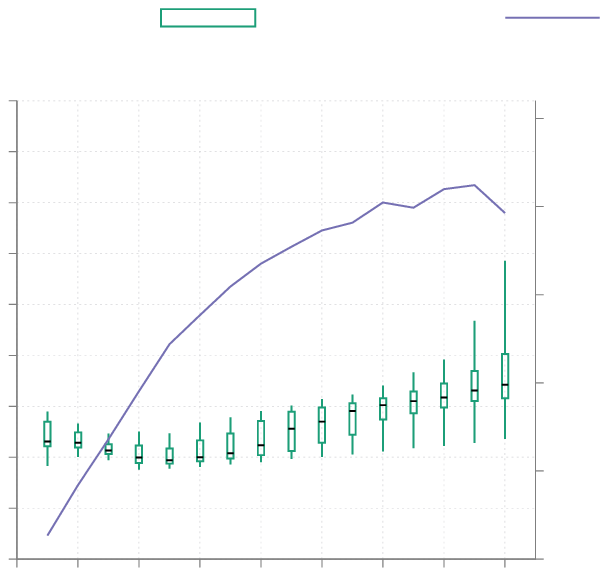}}
  \caption{\small SafeGENI authorization cost with delegations of length 6}
  \label{fig:safegeni-perf}
\end{subfigure}
\caption{\small Performance comparison of \keyword{SAFE} applications.
Subplots~\ref{fig:safedns-perf} and~\ref{fig:safegeni-perf} show the
performance of SafeNS and SafeGENI systems respectively.}
\label{fig:perf}
\end{figure*}

We evaluate \keyword{SAFE} on mix of cache configurations, micro-benchmarks,
and real applications. We seek to answer three questions:

\begin{itemize}
\item[Q1.] Does \keyword{SAFE} achieve acceptable performance? How
\keyword{SAFE} compares to ACLs and capability based access control where
policies are attached directly to entities rather than managed independently
through \keyword{SAFE} reference monitor? What is the overhead of invoking
\keyword{slog} through \keyword{slang}?
\item[Q2.] Are trust logics practical for use and deploy in real applications?
What is the programming effort required to build secure applications using
\keyword{SAFE} as the foundation for trust management and access control?
\item[Q3.] How does set linking and context caching improve performance across
space (cross-sharing of common \keyword{slogsets} among multiple simultaneous
queries) and time (caching frequently accessed \keyword{slogsets})?
\end{itemize}

All our authorizer experiments are conducted on eight core Intel Xeon CPU E5520
@ 2.27GHz processor with hyper-threading enabled and 8 GB of RAM running
CentOS 5.10. The SafeSets cluster consists of four VMs, each with a single core
Intel Xeon CPU E5620 @ 2.40GHz processor and 1GB of RAM interconnected by a 1Gb
network, all running Ubuntu 14.04. The authorizer access the SafeSets store
over WAN. We use unmodified Riak 2.0~\cite{riak} as our key-value store for
SafeSets guarded by \keyword{SAFE} as a proxy ``shim'' to authorize writes to a
\keyword{slogset}.

\subsection{Micro benchmarks}
To answer the first question, we use micro-benchmarks to evaluate
\keyword{SAFE} performance. Table~\ref{tab:ops} show the overhead of basic
operations in SAFE on a 1kB of payload per certificate. The \keyword{slogset}
identifiers use \emph{SHA-256} for hashing and base64 for encoding, which
result in fixed 44 byte strings. For signing, we use 2048-bit RSA keys. The
\emph{null} inference is the minimum latency penalty of querying \keyword{slog}
inference through \keyword{slang}. The overhead is proportional to the size of
proof context size and the number of environment variables which are globally
substituted from \keyword{slang} to \keyword{slog}. A fetch here is a single
\keyword{slogset} retrieved from SafeSets without traversing any links. A fetch
is verified for its authenticity---by verifying the signature on the
\keyword{slogset}---at the authorizer. On the contrary, posting a
\keyword{slogset} requires the client to sign its contents and send it to the
SafeSets server over the secure channel to prevent replay attacks. The SafeSets
server nodes are guarded by \keyword{SAFE} to determine whether the requesting
principal has write access to the \keyword{slogset} either directly or through
{\tt speaksFor*} capability. For the micro-benchmark, both fetch and post
including the cost of verification and signing the \keyword{slogset}.
Table~\ref{tab:ops} shows latency of post is 4X times the latency of fetch:
post is expensive since each post is idempotent and performs read-modify-update
to a \keyword{slogset}.

To analyze the cost of inference, we simulated delegation chains which the
number of unifications exactly matches the proof length. We measure the latency
by varying the length of number of unifications matching the goal from 1 to
1024 and the controlling the degree of backtracking, $b_d$. When $b_d$ is set
to zero, we have no backtracking and the length of the proof chain is linear in
terms of the unified goals in the input query. Setting $b_d$ to zero
approximates proof-carrying-authorization (PCA~\cite{Bauer02:pca}) where the
length of the input is exactly the length of the proof chain. In our
applications of SAFE, we observed that backtracking scenarios may occur with
attribute-based-delegation, where the principal on a goal is variable and the
proof context have multiple rule heads matching on the same goal. However, SAFE
uses indexing on multiple parameters which often reduces the $b_d$ to a small
value. Index optimization is a work in progress.  In our next experiment, we
set $b_d$ to 10, i.e., each goal can have at most 10 possible rule heads
matching with the goal and the length of the proof chain may grow exponentially
with the input size.  Figure~\ref{fig:proofLength} shows the latency
measurements of when varying proof chain length and $b_d$.  When $b_d$ is zero,
latency grows linearly with the proof length, which is expected.  However, when
$b_d$ is 10, it is interesting to note the latency remains linear and only
deviates from a linear scale at when the number of unifications matching the
goal (proof length in this case) is 600 or above.  The result shows that set
linking and tailoring proof contexts is important to keep $b_d$ low, which will
in turn help to scale the inference cost linearly with the size of the input.

The latency costs in Figure~\ref{fig:proofLength} show that SAFE inference
takes 0.1 ms per unification, which is competitive with respect to identity
based ACLs, which needs only one fact checked. Further, the plot shows that
comparison of inference costs when input is feed directly to a slog interpreter
vs. executed from the slang program. Recall that slang program will in turn
invoke slog interpreter after substituting all the environment variables. The
plot shows that overhead of calling slog from slang is minimal ($<5$\%).

\begin{table}
 \small
 \centering
 \begin{tabular}{p{2cm} r r} \\ \hline
      & \# Rules & \# SLOC \\ \hline \hline
    SafeSets & 2 & 15\\ \hline
    SafeNS & 7 & 40\\ \hline
    SafeGENI & 30 & 110\\ \hline
  \end{tabular}
\caption{\small Analysis of programming effort for building declarative trust
applications in \keyword{SAFE}.}
\label{tab:safesys-summary}
\end{table}

\begin{figure}[t!]
  \begin{center}
    \centering
    \scalebox{.9}{\input{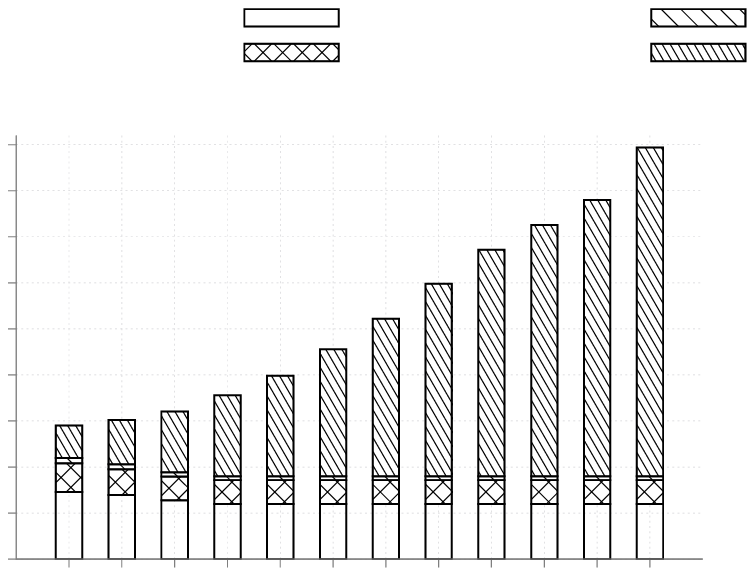}}
  \end{center}
  \caption{\small End-to-end authorization costs of SafeGENI with varying delegation
  lengths.}
  \label{fig:geniVaryingDelegation}
\end{figure}

\subsection{Applications of SAFE}
\label{sec:apps}
We built authorization systems for three practical applications in
\keyword{SAFE}. Table~\ref{tab:safesys-summary} shows the modest effort
required to build applications using \keyword{SAFE}. 

First, SafeSets uses \keyword{SAFE} as a proxy ``shim'' guarding write access
to \keyword{slogsets}. The post authorization for SafeSets involves validating
the speaker that signed the set on the server to determine the set ownership.
SafeSets is also a good example to illustrate the application of {\tt
speaksFor} and {\tt speaksForOn} predicates, which are implemented at
\keyword{slang} as discussed in~\ref{sec:slang}.

Figure~\ref{fig:safesets-direct-post} show the performance comparison of
SafeSets write (a write is a post with \keyword{slogset} signing exclued on the
client) without any authorization performed on the server, where as
Figure~\ref{fig:safesets-withoutVerify-post} show the performance comparison
without any authorization but \keyword{slogset} parsed by the server
(baseline).  Figure~\ref{fig:safesets-authorized-post} show the performance
comparison of SafeSets write with server authorization using \keyword{SAFE}.

The measurements show the peak throughput drops by 63\% due to parsing
overhead---our parser is not optimized---and the median latency increases by
50\% per write operation. If we compare the parsed \keyword{slogset} on the
server with the cost of authorization, then we observe that peak throughput
drops by 8\% and the median latency increases by 9\% per write operation. These
plots show that \keyword{slog} validation overhead is less than 10\% for simple
proofs that emulate ACLs.

Second, we implemented secure name service---SafeNS, discussed as
in~\ref{sec:safedns}. For delegations of length four (the average request
sub-domains for a DNS query is three), the 95\% latency of proof validation
cost of SafeNS is 6ms, which is a fraction of DNS lookup latency (in the order
of tens of ms). Figure~\ref{fig:safedns-perf} shows the latency and throughput
measurements of a proof validation for a NS query with delegation length four.
We achieved a throughput of 1600 auth ops/sec on our test suite with one
authorizer node.

Third, we prototyped authorization system for GENI, which is a networked
infrastructure-as-a-service (IaaS) system with autonomous IaaS providers linked
in a federated trust structure. GENI serves as a full-featured network trust
example that includes distributed objects (groups and slices\footnote{A slice
is a set of resources requested by an user}) with hybrid capability-based
access control, multiple object authorities, authority services for group
membership and federated identity management, and a common root trust anchor
that endorses the authorities and member sites.

The GENI trust architecture defines several classes of authority services to
manage user identity and authorize user activity. These services are
decentralized: each authority service may have multiple instances, and the set
of instances may change over time. In addition, users may delegate various
rights to one another using a capability model. SafeGENI specifies all of these
structures using logic and implemented in 110 SLOCs of \keyword{slang}.  For
delegations of length six, the 95\% latency of authorization cost for SafeGENI
is under 10ms with a throughput of 1400 authz ops/sec. The end-to-end
authorization cost including fetching from SafeSets, validating the sets and
ripping the crypto, building a context cache is under 20ms (see
Fig~\ref{fig:geniVaryingDelegation}). Most GENI delegations are smaller than
length 6. Figure~\ref{fig:safegeni-perf} shows the latency and throughput
measurements of authorization costs of SafeGENI.


\begin{figure}
  \begin{center}
    \centering
    \scalebox{.7}{\input{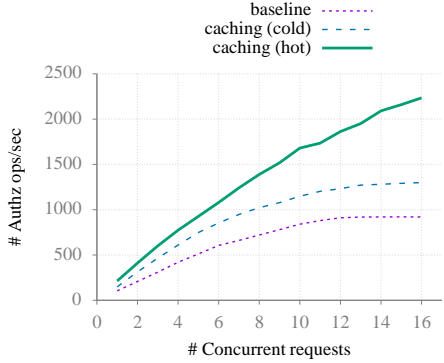}}
  \end{center}
  \caption{\small Throughput measurements for SafeGENI benchmark with $N$ set as 1024
  and $M$ set as 4096. The measurements show caching proof contexts can help to
  achieve linear scaling of throughput.}
  \label{fig:geni}
\end{figure}

\subsection{Impact of caching}
\label{sec:cache}

To measure the impact of caching and context linking on performance, we
benchmark SafeGENI using a standard GENI workload with a mix of $N$ users and
$M$ resources. We set up delegation chains so that any user is at most
$log_{2}(N)$ delegations away from accessing a resource. We measure the
throughput and latency of the mix under three scenarios: (i) Baseline case: all
certificates are processed in their entirety for each request with no caching
involved. This includes retrieving all certificates from in-memory, validate
crypto and speakers, render them to set cache, merge them to context cache, and
querying the inference.  (ii) Cold caching with monolithic contexts: the raw
certificates are cached in memory but the \keyword{slogset} cache and the
context cache are build on-demand. These context caches and \keyword{slogset}
caches are monolithic in that their life span tailored to a given request.
(iii) Hot caching with set references and proof context cache enabled. Here we
cache the proof contexts which enables shared credentials among queries is
readily available through the context cache.

Figure~\ref{fig:geni} shows the peak throughput measurements for the three
scenarios with $N$ set as 1024 and $M$ set as 4096. The throughput for baseline
case flattens out fast as expected since each request is processed in its
entirety, i.e., by validating the certificates, ripping the contents, and
evaluating the query. Caching the certificates improves throughput by 40\%.
However, the throughput flattens out after 10 concurrent requests. With hot
caching, the proof contexts are shared across the queries resulting shared
trust policies and credentials avoiding the re-rendering to proof context. The
measurements show that throughput scales linearly with hot contexts and
demonstrates the useful of caching proof contexts.

\section{Discussion}
\label{sec:discussion}
The \keyword{SAFE} approach to managing credentials also raises some potential
concerns.

{\bf Malicious content.}  Issuers may write malformed certificates to SafeSets
or generate a malformed credential DAG, e.g., by creating cycles in the DAG.
The SAFE fetch procedure rejects malformed certificates and detects cycles.
Valid certificates contain only slog statements, which share the termination
properties of pure datalog: all queries terminate.  However, issuers may create
very large or costly \keyword{slogsets} to mount a denial of service attack.   An
authorizer may bound the size of incoming logic sets and query contexts at its
discretion.

{\bf Accountability.}  Policy mobility relies on participants to enforce the
policy rules of others.   In general, entities control their own authorization
decisions and have power to do harm only to the extent that others trust them.
For example, a GENI aggregate that ignores policy conditions may be unduly
promiscuous with its own resources, but it cannot affect access to the
resources of others.  Moreover, all entities are strongly accountable (in the
sense of CATS~\cite{yumerefendi07strong}) for certificates they post to
SafeSets representing the result of access decisions.  Accountability is an
active research topic.


{\bf Confidentiality.}  Synthesized identifiers raise the issue of confidentiality
of policy rules and other logic material stored in SafeSets.  
If an entity wants to protect a confidential logic set it may salt the label:
it is infeasible to guess a hashed identifier that is effectively random.  We
emphasize that the protection for writing to SafeSets is stronger: a client
must possess a principal's private key in order to write to a logic set that
the principal controls.

{\bf Reclamation.}  Logic material may accumulate in SafeSets over time.
SafeSets may delete any set after it has expired: all slogsets have expiration
times. Even so, issuers may use unreasonable expiration times or simply post
useless data to the store. SafeSets authenticates each issuer by its public
key, but quotas are of no help if an issuer can mint new keys at will. One
option is to apply a SAFE access check for posting.  Another option is to
arrange the store so that each issuer provides and manages its own storage
(e.g., via a Web server).

{\bf SafeSets failure.}  Managing SafeSets as a decentralized key-value store
can be a scalable and reliable solution. One or more entities may control the
SafeSets servers. A faulty or malicious server can destroy content or block
access to it. However, it cannot subvert the integrity of the system because
all logic sets are signed by their issuers.

\vspace{-2mm}
\section{Related Work}
\vspace{-2mm}
As flexible and extensible trust logic for federated network systems, SAFE
build upon a wealth of prior results. Thus, for SAFE, the closely related work
covers the entire fields of study---including authorization
logic~\cite{Becker10:secpal,sirer11:attestation,detreville:binder:2002,jim:sd3:2001,Bauer02:pca,Li03:delegationlogic},
trust management~\cite{blaze03:keynote, blaze96:trust}, proof-carrying
authorization~\cite{appel99:proof-carrying, Lesniewski-Laas07:alpaca}, and
scalable storage systems~\cite{DeCandia07:dynamo}.

In general, trust logics apply common axioms of ABLP access control
calculus~\cite{abadi93:accesscalculus}. In particular, every entity {\it
controls} its own beliefs through   \comment{ An entity may state an inference
rule for its own beliefs: a rule with a tag ``$A$ {\it says}'' in the head is
valid only if it is an authenticated statement made by $A$.  If $A$ does make
such a statement, then other entities may use the rule to infer $A$'s beliefs
based on a given set of ground facts.   These} axioms are known as {\it
Hand-off} and {\it Bind} respectively~\cite{abadi08}. They enable sound trust
policies and mobility of policy rules.

Many of these logics offer features that are not present in slog: our research
goal is not to advance trust logics, but to facilitate their practical use.
While at present we see little need for features such as threshold/manifold
structures or negation, slog could grow to incorporate them without
compromising tractability, following SecPAL~\cite{Becker10:secpal} and
NAL~\cite{schneider11:nal}.  However, SAFE supports a simple negation
restricting only queries to contain {\tt not} predicate to allow {\tt deny}
conditions (blacklists).

Some logics use (e.g., {\tt speaksFor} and {\tt speaksForOn}) as primitive
axioms to delegate trust~\cite{abadi93:accesscalculus, Howell:2000,
schneider11:nal}. On the contrary, SAFE represents delegation with rules in
datalog-with-{\tt says}, and uses {\tt speaksFor} predicates only for joint
ownership or to authorize third-party attributions, e.g., for service proxies
or portals that are trusted to issue statements for which the named speaker is
another party. 

The paper uses set linking and linked contexts to address the challenge of
assembling the context for a proof.  One option is to require the caller to
submit the proof to the authorizer, as in Proof-Carrying
Authorization~\cite{Bauer02:pca}. PCA merely shifts the burden of assembling
the context to the caller.  Our premise is that PCA is unnecessary for a simple
trust logic like datalog-with-{\tt says} and careful context management:
constructing a datalog proof from a small proof context can be fast.  In
contrast, the AF logic that underlies PCA is intractable in the general case.
The other option is bearer credentials that are only verified by the specific
\emph{target} service rather than any authorizer as proposed by
Macaroons~\cite{macaroons:2014} using HMACs.

More fundamentally, whether or not PCA is used, the caller must identify the
relevant credentials to send with a request, or else the authorizer must obtain
them by some other means. Previous approaches to credential discovery are based
on a distributed query model
(e.g.,~\cite{li01:discovery,Bauer05:DPA,bauer07:proving}).  In SPKI/SDSI, the
name resolution~\cite{clarke01} requires the authorizer to resolve the relevant
certificates among a potentially large set of certificates in the \emph{right}
order. On the contrary, explicit set linking makes credential discovery
scalable and practical. Many previous works ignore the problem and presume that
the caller will identify the correct credentials and pass them in the request.
In these systems, the receiver/authorizer validates and checks the request
credentials even if it has already received them for a previous request. In
contrast, linked context sets naturally support caching and pass-by-reference
for credentials.

SafeSets facilitates certificate sharing via a highly available shared store.
The early X.500 model proposed a distributed certificate repository, as
discussed by the SPKI/SDSI authors~\cite{spki-rfc2693}, who judge it to be
impractical. A key difference is that SafeSets links certificates by set
identifiers, and does not rely on global principal names other than public
keys, as advocated by SPKI/SDSI. Various other systems have proposed
certificate storage for use in credentials-based authorization
(ConChord~\cite{Ajmani02:ConChord}, CERTDIST~\cite{Sevinc09:CERTDIST}) using
various indexing and naming schemes, but they do not support set linking.
SafeSets supports secure unforgeable set identifiers for a key/value store by
qualifying them with a public key hash.  A similar technique has been used in
many DHT applications~\cite{SFS:SOSP:1999,SUNDR:2004}.

\label{sec:related}

\vspace{-2mm}
\section{Conclusion}
\vspace{-2mm}
\label{sec:concl}
\keyword{SAFE} is a open-source trust management system~\cite{safe-repo} that
uses a declarative trust logic to represent policies, endorsements, and
delegations.  What is novel about \keyword{SAFE} is the integration of the
trust logic with a scripting language (``slang'') and shared storage
abstraction for authenticated logic content.  These elements work together to
simplify and automate many aspects of networked trust. In the implementation
for this paper we materialize logic sets as signed certificates, as in
SPKI/SDSI, and store them in a scalable certificate store called SafeSets. Each
stored certificate is named by an identifier suitable for indexing and caching
linked certificates or logic sets.

We use this combination to address three fundamental problems: how to identify
the content that is relevant to a given trust decision, how to manage the flow
of credentials through the system including caching, and how to incorporate
updates to outstanding certificates. Experience with SafeNS and SafeGENI shows
that the approach is practical for a complex network trust system.

\vspace{-2mm}
\section{Acknowledgemnts}
\vspace{-2mm}
The first author is partially supported by NSF grants OCI-1032873 and
CNS-1330659 for this work. We thank Jon Howell for his feedback on the paper
and pointer to the Snowflake project. We also thank Mike Reiter for helpful
comments.

\vspace{-3mm}
\bibliographystyle{abbrv}
{
\bibliography{main}
\vspace{-2mm}
}
\end{document}